# On the Simulation of Enzymatic Digest Patterns

Jonathan J. Hunt [1], Randall G. Cameron [2] and Martin A.K.Williams [1], *

[1]*Institute of Fundamental Sciences, Massey University, Palmerston North, New Zealand.*

[2]*USDA, Citrus and Subtropical Products Laboratory, 600 Ave S, N.W. Winter Haven, Florida 33881, USA*

**Abstract**

A simulation methodology for predicting the time-course of enzymatic digestions is described. The model is based solely on the enzyme's sub-site architecture and concomitant binding energies. This allows sub-site binding energies to be used to predict the evolution of the relative amounts of different products during the digestion of arbitrary mixtures of oligomeric or polymeric substrates. The methodology has been specifically demonstrated by studying the fragmentation of a population of oligogalacturonides of varying degrees of polymerization, when digested by endo-polygalacturonse II (endo-PG II) from *Aspergillus niger*.

* Corresponding author, m.williams@massey.ac.nz

## 1. Introduction

The importance of degradative enzymes across the whole of Nature can hardly be over-emphasised. All biopolymers are, at some point, disassembled during recycling or digestion. Additionally, the routine remodelling of biopolymer fine structures in order to maximise their *in-vivo* functionality often involves carrying out modifications to their degree of polymerisation (DP). Furthermore, coupling a thorough knowledge of the action of degrading enzymes with the measurement of the structures of released fragments has the potential to yield information on the fine structure of pre-digested substrates. Such an approach still holds particular promise for polysaccharide systems, where the molecular biology tools that have so advanced nucleotide and protein sequencing are simply not available. Hence, polysaccharide degrading enzymes are not only of fundamental biochemical interest but also, owing to the substrate sequence sensitivity of their enzyme-substrate interactions, offer the hope of developing a tangible link between the structures of released digest fragments and polymeric fine structure [1-9].

Models of polysaccharide-enzyme interactions routinely begin with the protein being described by a series of sub-sites, each capable of binding a sugar residue [10]. These sub-sites consist of groups of amino-acids, with their chemical nature giving rise to differential affinities between such sites and the substrate. Sub-sites labelled +1 and -1 indicate the position of the active site, while surrounding sub-sites, extending from the active site to the proximities of the binding cleft, play a role in binding the substrate [11]. By observing the preferred cutting position of substrates that are smaller than the number of sub-sites in the enzyme binding site, relative affinities and hence binding energies of different sub-sites can be obtained [12-17]. Such maps can be used to give support to molecular models of the enzyme-substrate interaction by providing experimental values that can be compared with chemical intuition based on the nature of the moieties within the amino acids of each designated sub-site. Once the relative importance of hydrophobic and charge interactions is inferred the effects of solution properties might also be rationalised. Such experiments also provide data that can be compared with attempts to directly model the enzyme-substrate interaction [18]. However, despite the many positive aspects of sub-site mapping procedures such results have not, as far as we are aware, been used to predict the general time-course of the enzyme digestion of arbitrary mixtures of oligomeric or polymeric substrates.

The study of the enzymatic digestion of biopolymer substrates has also been a fertile area of research in its own right. In particular, experimental techniques for obtaining data on biopolymer digest fragments and studying the structure of complexes continue to improve [19-23]. Modelling approaches meanwhile have focussed largely on solving differential equations, [24] developing recursion schemes [25] and using methods that assume statistical models of chain structure and a strict set of prescribed enzymatic rules [8]. In this work we show that a simulation of substrate digestion, derived simply from the enzyme's sub-site architecture and attendant binding energies, can successfully describe the experimental time-course of digestion of distributions of substrate molecules of varying lengths. We demonstrate this specifically by taking a well defined starting mixture consisting of substrates of various DPs, measuring the evolution of the concentration of all species during enzymatic processing, and comparing the results with a simulation. Further, we examine the sensitivity of the form of the predicted digest pattern to variations in the binding energies associated with the different sub-sites.

Oligogalacturonide substrates digested with an important pectin degrading enzyme, endo-polygalacturonse II (endo-PG II) from *Aspergillus niger*, has been selected as a model system, so that the generic digestion problem can be discussed while producing specific results of importance for many areas of plant science. The ubiquitous occurrence of the polysaccharide pectin in the cell walls of land plants, taken together with the large number of pectin-modifying enzymes encoded in the genomes of plants [26] and their pathogens, [27] clearly points to the importance of pectin, and pectin derived compounds, in diverse aspects of plant physiology. Indeed, the products of pectin digestion, the oligogalacturonides, play a key role in triggering plant defence [28]. Endo-PG II has been classified as a family 28 glycosyl hydrolase and proceeds by an inverting mechanism [29, 30]. Its preferred substrate is galacturonic acid but it can also digest partially methylesterified galacturonides and co-polymers of galacturonic acid and its methylesterified counterpart (pectin), although there is an absolute requirement for sub-sites -1 and +1, which straddle the scission point, to be unesterified [31]. Its sequence and crystal structure have been determined and a number of mutants have been studied [32, 33-35]. Bond cleavage frequencies have also been obtained for the degradation of galacturonic acid oligomers up to the hexamer, allowing sub-site binding energies to be mapped [31]. In addition, while tetramers and trimers each have only one productive binding mode that spans the active site, the relative kinetics of fragmentation [36] can provide differential binding energy estimates and hence probe the affinity of the extra sub-site involved in tetramer degradation. Fig. 1 shows the current best model of the endo-PG II sub-site architecture.

## 2. Materials and Methods

### 2.1 Oligomeric Substrate

A set of oligogalacturonides containing degrees of polymerization between 2 and 17 was generated by selective precipitation of a partially digested polygalacturonic acid (PGA) as described previously [37]. Briefly, a 2% (w/v) solution of the free acid of PGA in 50 mM lithium acetate at pH 4.7 was digested with 0.05 U mL$^{-1}$ EPG (Lot 00801, Megazyme International Ireland Limited, Bray, Ireland) for 4.5 hrs at room temperature with constant stirring. The pH of the digested LiPGA was lowered to 2.0 with concentrated HCl and stored overnight at 4 °C. The precipitate was pelleted by centrifugation at 23,500 g for 30 min at 4

°C. The pelleted precipitate, representing high DP oligomers was removed. The supernatant, containing the low- and medium-DP oligomers, was brought to 50 mM sodium acetate (NaOAc) and 22.5% EtOH and then placed at 4 °C overnight to precipitate the medium-DP fragments (DP 8-24). Following centrifugation as described above, the supernatant was decanted. The pelleted material was solubilized in 50 mM LiOAc and then re-precipitated by adjusting the solution to 50 mM NaOAc and 22.5% EtOH. This material, representing the medium-DP oligomers, was centrifuged again and the pellets were solubilized in 50 mM LiOAc and stored at 4 °C. Initial characterisation was carried out using HPAEC with an evaporative light scattering (ELS) detector [38,39] and subsequently by capillary electrophoresis using UV detection [19, 22].

**2.2 Enzymes**

Endo-PG II (EC 3.2.1.15) from *Aspergillus niger* was prepared as described previously [40]. Digests were carried out by incubating 1.0 ml of the oligogalacturonide mixture, at a total concentration of 3 % galacturonic acid, and pH 4.2, with 20 μL of the enzyme solution, that was in turn generated by diluting 25 μL of a 7.5 mg ml$^{-1}$ protein stock into 2.0 mL of 50 mM acetate buffer at pH 4.2. All experiments were carried out at (30±1) °C by keeping the digest mixture in a waterbath. At various times aliquots were removed, the enzyme denatured by rapid heating to 95 °C, and the current concentrations of the various oligomeric species recorded using CE. Thus a picture of the time-course of the digestion was recorded.

**2.3 Capillary Electrophoresis**

Experiments to separate, identify and quantify oligogalacturonides of varying degrees of polymerisation were carried out using an automated CE system (HP 3D), equipped with a diode array detector. Electrophoresis was carried out in a fused silica capillary of internal diameter 50 μm and a total length of 46.5 cm (40 cm from inlet to detector). The capillary incorporated an extended light-path detection window (150 μm) and was thermostatically controlled at 25 °C. Phosphate buffer at pH 7.0 was used as a CE background electrolyte (BGE) and was prepared by mixing 0.2M $Na_2HPO_4$ and 0.2M $NaH_2PO_4$ in appropriate ratios and subsequently reducing the ionic strength to 90 mM. At pH 7.0 galacturonic acid residues are fully charged and while the oligomers are susceptible to base-catalysed β-elimination above pH 4.5, no problems were encountered during the CE runs of some 20 minutes at room temperature. All new capillaries were conditioned by rinsing for 30 minutes with 1 M NaOH, 30 minutes with a 0.1 M NaOH solution, 15 minutes with water and 30 minutes with BGE. It was found that for the samples used in this study similar harsh washing of the capillary was also required between runs. Detection was carried out using UV absorbance at 191 nm with a bandwidth of 2 nm. Samples were loaded hydrodynamically (various injection times at 5000 Pa, typically giving injection volumes of the order of 10 nL), and typically electrophoresed across a potential difference of 20 kV. All experiments were carried out at normal polarity (inlet anodic) unless otherwise stated. Samples of mono-, di-, and tri-galacturonic acid, used as standards, were obtained from Sigma-Aldrich Corp., St.Lois, MO, USA.

## 2.4 Simulation

Oligomeric distributions of starting material were modelled by a simple one dimensional array in which the elements were assigned to a character, L, M or R, denoting the left (non-reducing end), middle, and right (reducing end) of chains. Enzymatic encounters were presumed to occur at random. Each encounter consisted of selecting a substrate binding position for sub-site -1 and summing the binding energies for all filled sub-sites in this position. It was assumed that each sub-site binding energy was unaffected by the presence or not of substrate at other sites. This summed energy was then used to generate a Boltzman factor that determined the probability of successful binding. This conversion to probability included a normalisation of the binding energy to that of 7 filled sub-sites (the accepted extent of the endo-PG sub-site architecture, as shown in Fig.1), so that such a contact always results in a successful binding event. Whether the binding was successful or not was determined by comparing this probability with the output of random number generator from GNU Scientific Library [41]. Using this random number generator ensured that the frequency of low-probability events was simulated accurately. In successful binding cases, if sub-sites +1 and -1 were both covered, then a scission was made and the array elements reassigned accordingly in order to denote the presence of new chain ends. After a selected number of iterations the array was interrogated to produce a distribution of fragment lengths, and it was this simulated data that was compared with the experimentally measured time-course. The simulation was carried out on an Intel P4 computer and takes only seconds to complete a simulated digest evolution starting with $10^4$ substrate molecules. The code was written in C++, compiled with Intel C Compiler (Version 9), and was executed using the scripting language Python within a Linux environment. Initially, solely tetramers, pentamers, or hexamers were fed to the enzyme model to ensure that the relevant number of different product possibilities (i.e. the known bond cleavage frequencies) were reproduced.

## 3. Results and Discussion

Fig. 2 shows the quantification of the starting oligogalacturonide distribution as measured by both HPAEC with ELS detection and CE using UV absorption. The agreement between these two quite different methodologies is remarkably good and provides further confidence in both techniques. The CE data was processed in order to quantify the amount of different oligomers present, as described in detail elsewhere [22], ensuring that the peak area was divided by the migration time in order to account for the different time that the separated species spend travelling past the detection window [42]. The peaks were unequivocally identified by spiking with commercially available samples of dimer and trimer. CE has been used previously in the study of oligogalacturonides and it is known that above a certain molecular length the hydrodynamic friction and charge scale symmetrically with the introduction of further sugar residues, leading to a loss of resolution. While the exact DP at which this occurs and the modelling of the mobility forms part of ongoing work it has previously been predicted to occur at around DP 15-20 [43]. Bearing this in mind the current oligomer mixture was selected as the starting substrate in order to ensure it would be possible to record the time-course of all starting oligomers.

The initial input for the enzymatic fragmentation model was the relative number of molecules of each oligomer, generated from the average of the HPAEC data and three CE datasets. The only other parameters used are a best estimate set of the binding energies for the endo-PG II subsites, shown in Fig. 1. Preliminarily, the number of starting substrate chains that could be used in order for the simulation to generate reproducible results was determined. For a chosen initial number of chains a digest simulation was run for fixed number of time-steps, the value of which was selected based on preliminary experimental data, and the resulting pattern recorded. This was then repeated with the same number of chains so that 10 estimates of the resulting pattern (the relative number of the different DP oligomers existing at a particular time) were obtained. From these a mean pattern was obtained, and the variance, $\sigma^2$, of the 10 repeats around this mean determined. This process was repeated for sets of solutions carried out with different numbers of starting chains, (keeping the number of iterations per chain constant), and the results are shown in Fig. 3. It is clear that the reproducibility of the calculation becomes better as more chains are used, as expected. An optimal starting value, taking into account the extra time required to perform the calculations for a greater number of chains, was deemed to be $10^4$ substrate molecules, and this was used for all subsequent simulations.

Experimental data were obtained for the evolution of the enzyme digest pattern at 0, 2, 4, 6, 8, 10, 15, 30, 45, 266, and 13000 minutes as described in the experimental section. A simulation of the digest process was also run. The simulated digest pattern data was typically stored after each additional 200 iterations. The calculation was terminated after $6*10^6$ iterations at which time the changes in the digest pattern consisted solely of trimer being slowly fragmented to monomer and dimer, with a further $10^6$ iterations generating concentration changes of a fraction of a percent. It is worth commenting at this point that it is experimentally observed that the trimer is indeed digested to the dimer and monomer but at a rate some 20 times slower than the tetramer degradation [36]. This rate is sufficiently slow as to essentially decouple the trimer processing from the rest of the digestion and therefore, in many previous studies, the relatively long-lived situation with a similar ratio of monomer, dimer and trimer has been taken practically to be the end of the digestion. In the model described here these two regimes emerge naturally as a consequence of the differential binding energies of the species of various lengths.

Initially the experimental and simulated digests were compared by eye in order to assess whether there were calculated digest pattern solutions during the simulated fragmentation process that manifested the experimentally found ratios of different DP oligomers. Encouraged by a reasonable agreement in the general time-course of the digest, a simple searching algorithm was subsequently written that took each experimentally determined digest pattern and searched through the calculated data in order to find the particular simulation (corresponding to a certain number of iterations) that best matched, as determined by a simple minimisation of the standard deviation of the experimental and simulated datasets. As a prelude to the formation of the standard deviation the patterns concerned were both normalised so that the total amount of galacturonic acid was the same in both cases. A scaling of the experimental data total amount was permitted during the pattern matching to allow for the possibility that not all the material was detected, but was found to not be significantly different from one when the matches were made.

Fig. 4 shows the comparison of the experimental and calculated digest patterns for these best fits. In general there is excellent agreement, particularly when experimental uncertainties in the relative amounts are taken into account. It should also be born in mind that while it is the number of molecules that is represented in these plots it is the absorbance that is actually detected in the CE experiments reported herein, which is proportional to the degree of polymerisation (with a minor exception for the monomer, as described previously elsewhere [22]). This means that small amounts of the dimer and monomer, as predicted at the beginning of the simulation are difficult to detect experimentally. Nevertheless, the agreement is good. It is interesting to use the correspondence of the simulated and experimental data to map the iteration number onto real-time, as shown in Fig. 5. The plot can be seen to show an excellent linear relationship up to 45 minutes, implying that despite the relatively simplistic nature of the model including the random nature of the encounters and the lack of dynamics and molecular detail, nevertheless, the previously measured sub-site binding energies do describe the time dependence of the real reaction well. The last two points were more difficult to match to a specific time owing to the slowness of the evolution, but simulations terminated at the number of iterations that correspond to such long times according to the data in Fig. 5 were indeed found to be consistent with the experimental data.

While in detail the value of the gradient of such a plot will depend, among other things, on enzyme and substrate concentrations it could be argued that performing one time-step in the simulation might be thought of as modelling a substrate-enzyme contact. The mapping of real-time onto equivalent simulation time-steps then, as shown in Fig. 5 suggests that there is roughly one cut attempt every 6 ms, that is; there is an encounter between enzyme and substrate on the order of milliseconds. A simple order of magnitude calculation of the collision frequency can be made by distributing the enzyme molecules homogeneously throughout the solution and calculating the time it would take the oligogalacturonides to move between them (the diffusion coefficient of the protein is expected to be at least ten times less than the sugar oligomers). Taking the enzyme concentration as 100 nM and an average substrate diffusion coefficient of $10^{-10}$ $ms^{-2}$ gives an average encounter frequency in the tenths of milliseconds regime that agrees reasonably for such a simple model.

We have shown then that the binding energies inferred from bond cleavage and kinetic data on small substrates (DP<7) can be utilised in a physically realistic model in order to reproduce the temporal evolution of the digest patterns of arbitrary mixtures of considerably longer starting substrates. This implies that, for this system at least, the binding energy of the sub-sites is not significantly modified by the presence or not of substrate in neighbouring binding sites, and further, that they are not significantly altered by any straining of the substrate.

Further simulations have also been carried out in order to assess the sensitivity of the predicted digest patterns to the values of the binding energies and hence address the question - how much variation could you have in the binding energy values before you obviously modify the form of the simulation and cannot match the experiments? The basic philosophy of these was the following. Twelve "times" were chosen and at each of these a digest pattern was simulated with the base, best estimate values of the endo-PG II sub-site energies as given in Fig. 1. These "times" were in fact the number of time-steps that gave digest patterns in the simulation that corresponded to experimental data recorded, some of which was shown in Fig 4. Subsequently the binding energies were altered in a known manner and the simulation was re-run, this time calculating the digest pattern at each 200 iterations. A searching algorithm then found the best match pattern generated by the simulation using the modified binding energies to the base case. The difference between this digest pattern, generated with modified energy variants and that calculated with the best binding energies, was then quantified by means of a simple variance, as described above.

Fig. 6 shows the results for a representative number of these calculations. Owing to the complexity of the combinatorial space of five binding sites, a simple approach was taken in which the binding energy of each sub-site was changed in isolation, between the limits of the original value minus 50 % to the original value plus 50 %, at 1 % increments. It can be seen that at very short times there is, as expected, little change in the calculated digest pattern even with large differences in the binding energies, simply reflecting the fact that there has been little change in the original distribution at this point. However, by six minutes it is soon apparent that changing the binding energies of sub-sites +2, -4 and -5 leads to a substantial change in the digest pattern. In order to generate a significant change however, alterations in the binding energies of the relevant sub-sites need to be of the order of 10-20 %. It is also clear that sub-sites -2 and -3 do not play a significant role in determining the form of the digest pattern in this region, as expected, at a time when no degradation of tetramers and trimers will have occurred. This trend essentially continues for the simulations run to a number of time-steps equivalent to 10 minutes. Sill later in the time-course of the simulated digest, by 30 minutes, it becomes apparent that there has been such a substantial time period involving the fragmentation of the smaller species that binding energy value of sub-site -5 becomes less important than +2 and -4 in terms of generating the final pattern. Towards the end of the digestion process the fragmentation pattern can be seen additionally to be sensitive to increases in the affinity at site -3.

In addition to carrying out these simulations, performed in order to show that the calculated form of the digest pattern is sensitive to the input values used for the binding energies, the ability to detect variances of the predicted magnitudes experimentally has also been examined. Fig. 7 shows the variances of predicted digest patterns, calculated with different binding energy variants, from that experimentally observed at 10 minutes. The plot clearly shows that if the binding energies of sites -5, -4 and +2 were modified by more than around 10% then the match *to the experimental data* would be detectably worse. In conclusion then, the fact that the simulated enzymatic fragmentation calculated based on sub-site binding energies matches experimental data is indeed significant; changes in the affinities of individual binding sites of greater than 10-20% would have produced detectable discrepancies in the ability of the simulated data to successfully match the experimental time-course.

## 4. Conclusion

A model for predicting the time-course of enzymatic digestions of arbitrary mixtures of oligomeric or polymeric substrates has been described. The success of such a simulation has been demonstrated by modeling the fragmentation of a population of oligogalacturonides of varying degrees of polymerization, when digested by endo-polygalacturonse II (endo-PG II) from *Aspergillus niger*. This allows sub-site binding energies obtained from the results of bond cleavage and kinetic experiments, carried out on oligomers with lengths less than the number of sub-sites in the binding cleft, to be used to predict the evolution of the relative amounts of different products during the digestion of mixtures of arbitrarily long oligomeric or polymeric substrates. The different phases of digestion within the fragmentation process naturally evolve from the differential binding energies of different DP substrates. This simple homo-polymeric case serves to demonstrate the promise of the technique and current work is involved in extending the modeling to the fragmentation of co-polymers, in which the different residue types can have different binding energies to the enzyme sub-sites.

Fig. 1. The current best model of the endo-PG II sub-site architecture, showing the position of the active site and the binding energies associated with the different sub-sites, that have been used in this work. Sub-site binding energies for sites -5, -4 and +2 have been taken from studies of bond cleavage frequencies; that for -3 from the relative kinetics of tetramer and trimer degradation; and that for -2 arbitrarily selected.

Fig. 2. The quantification of the starting oligogalacturonide distribution as measured by both HPAEC and CE.

Fig. 3. The variance within sets of 10 repeat simulations carried out with different numbers of starting chains. The inset shows the data on a logarithmic axle.

Fig. 4. The comparison of the experimental and calculated digest patterns at selected times during the fragmentation of a starting oligogalacturonide solution with endo-PG II.

Fig. 5. The number of iterations required in the simulation in order to generate a simulated digest pattern that best matches that found experimentally at the corresponding time.

Fig. 6. The variance between calculated digest patterns generated i) with the best estimate sub-site binding energies and ii) with the indicated sub-site energy specifically modified in isolation, between the limits of the original value minus 50 % to the original value plus 50 %, at 1 % increments. (a) 2 minutes, (b) 10 minutes, (c) 30 minutes, 13000 minutes.

Fig. 7. The variances of predicted digest patterns, calculated with different binding energy variants, from that experimentally observed at 10 minutes. The plot clearly shows that if the binding energies of sites -5, -4 and +2 were modified by more than around 10% then the match *to the experimental data* would be detectably worse.

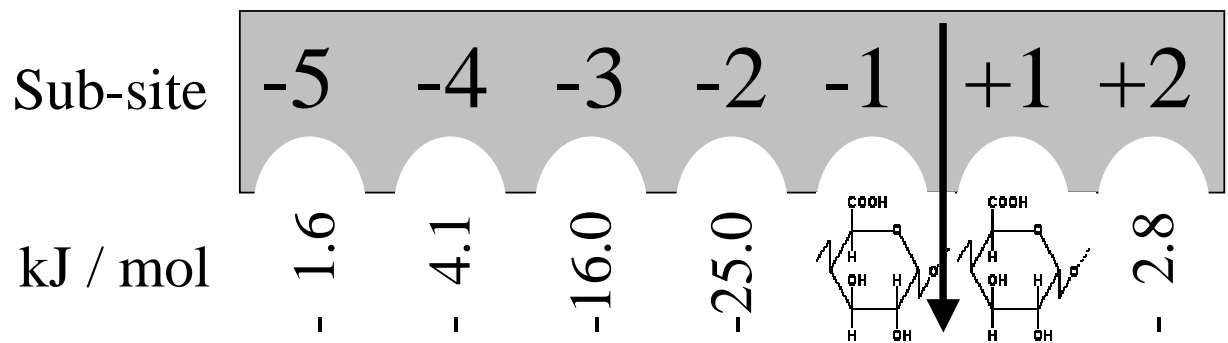

FIGURE 2

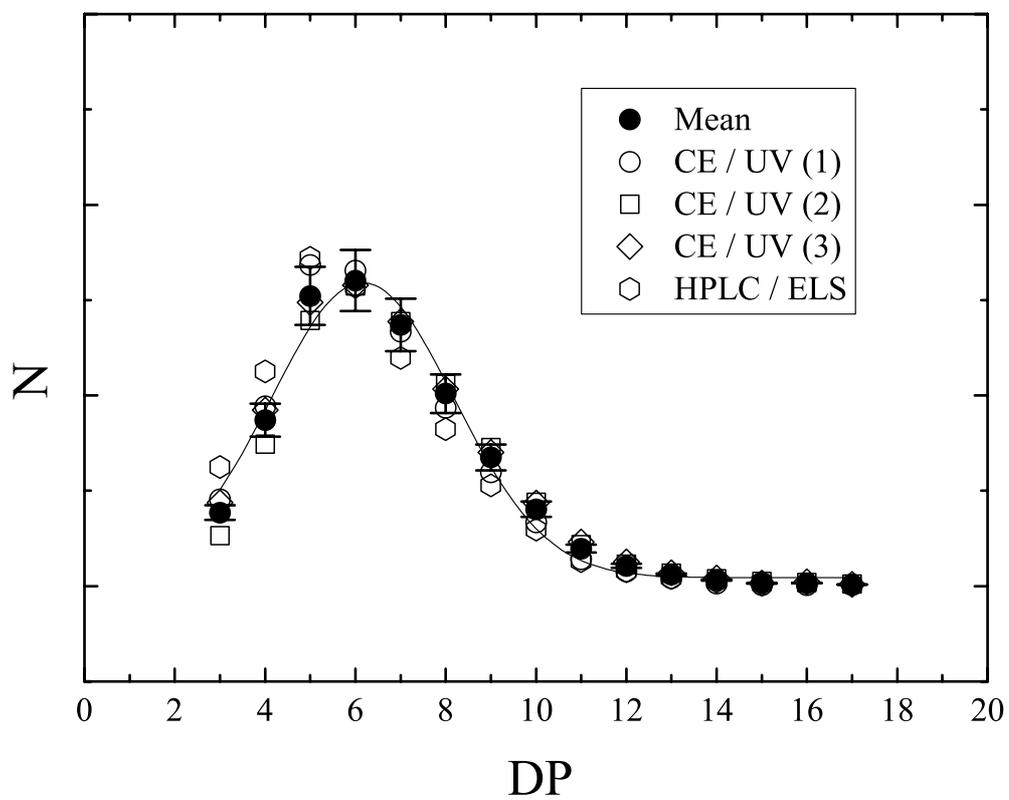

FIGURE 3

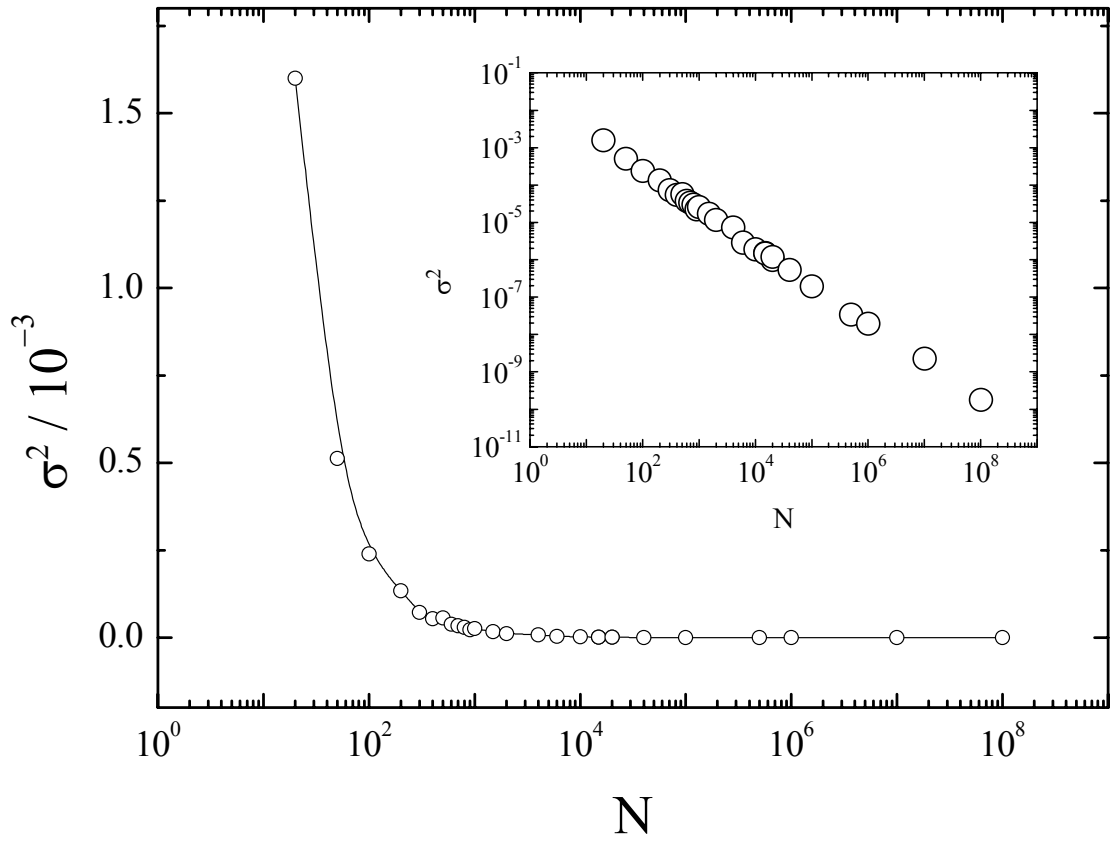

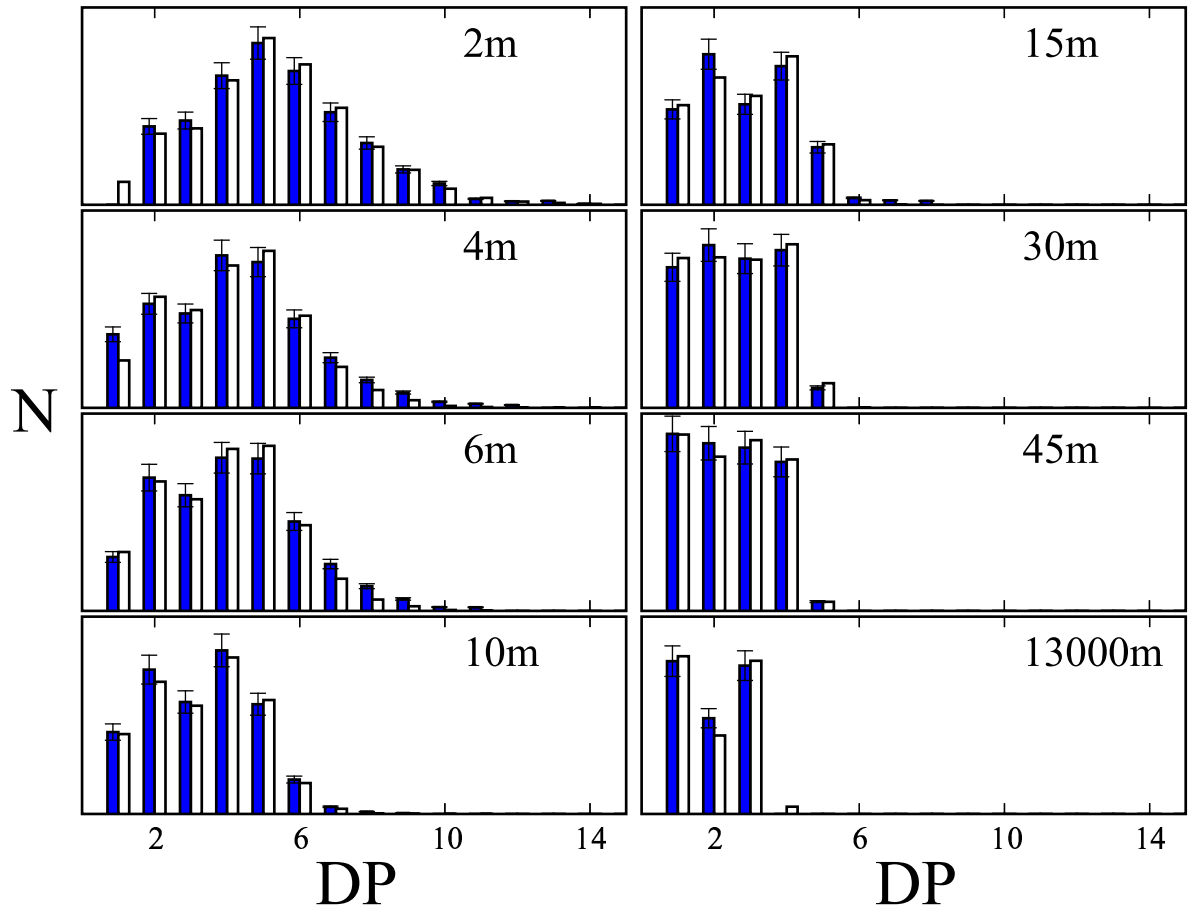

FIGURE 5

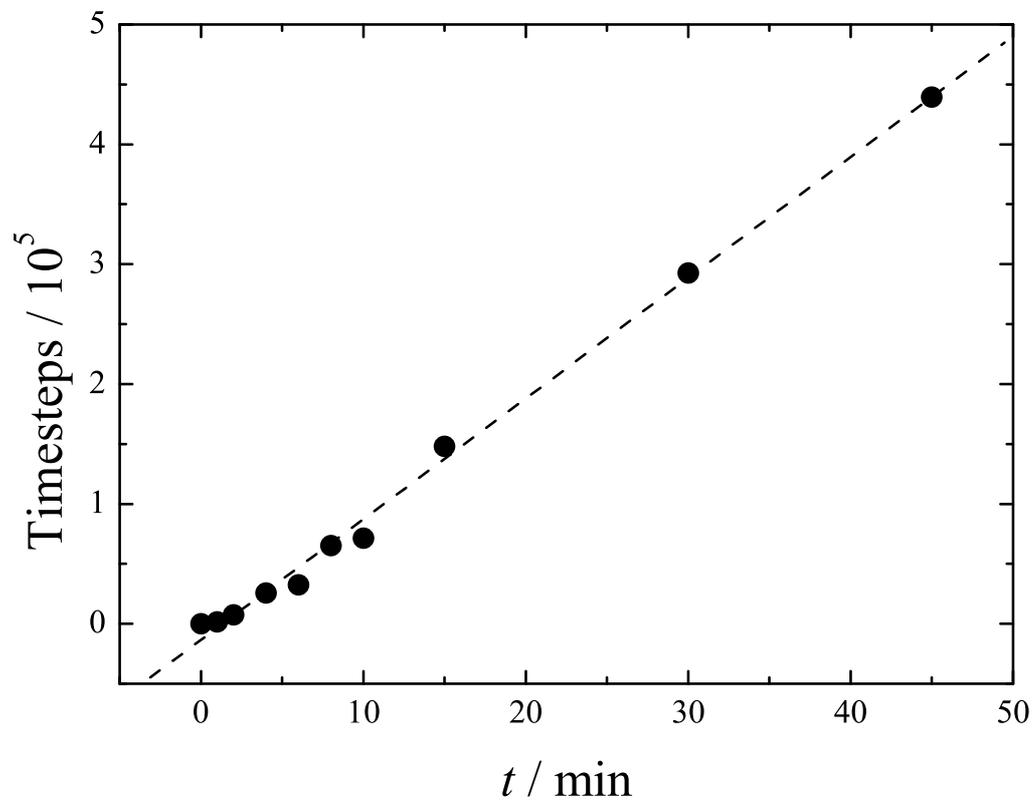

FIGURE 6

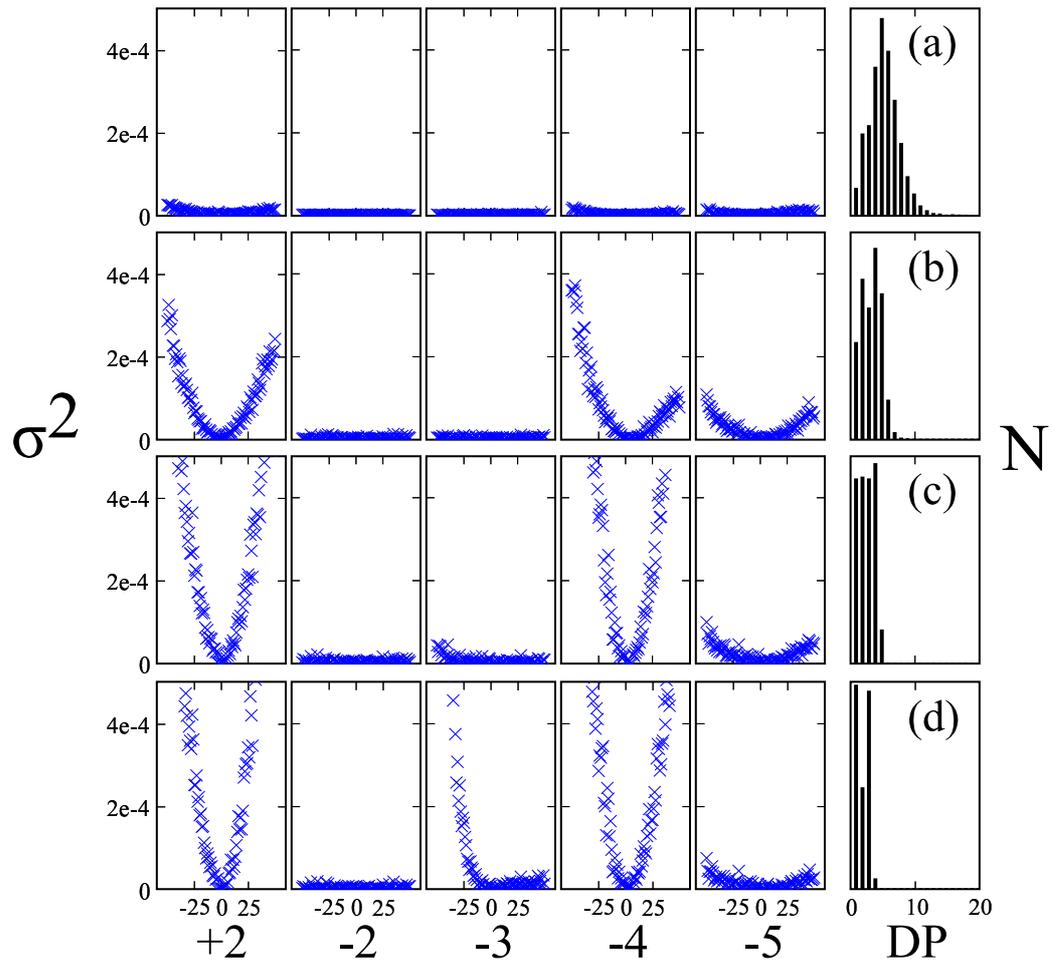

FIGURE 7

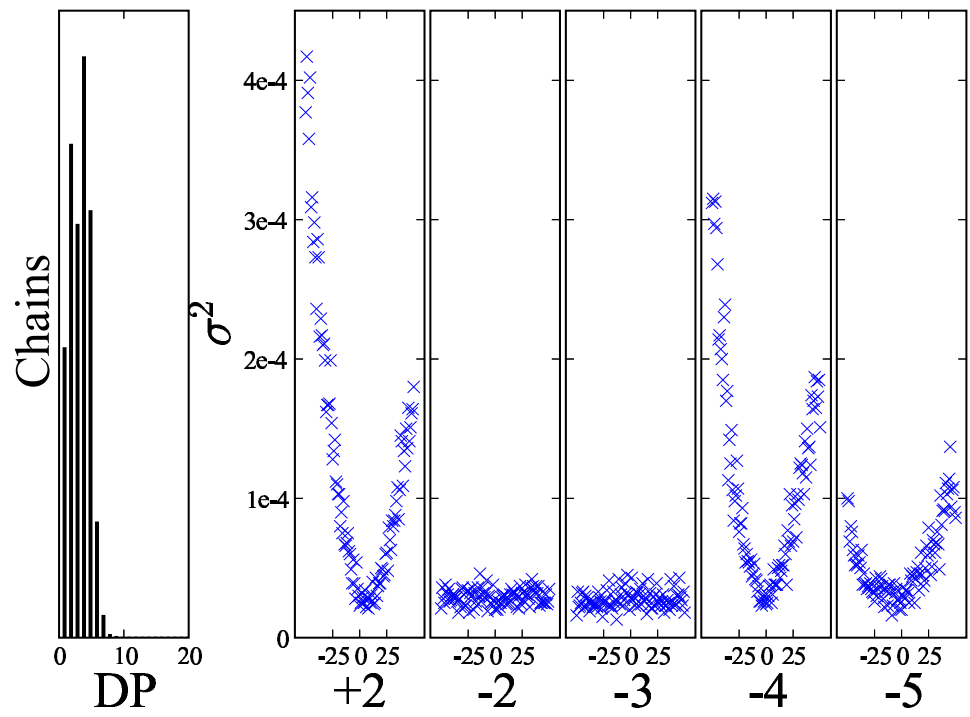